\documentclass[onecolumn]{aastex631}

\usepackage{graphicx}	
\usepackage{amsmath}	
\usepackage{subfigure}
\usepackage{natbib}
\usepackage{color}
\usepackage{tabularx}
\usepackage{longtable}
\usepackage{bm}
\usepackage{threeparttable}
\usepackage{enumerate}
\usepackage[normalem]{ulem}
\usepackage{verbatim}

\newcommand{\code}[1]{{\texttt{#1}}}

\begin{document}

\title{Gaia Astrometry and MIKE+PFS Doppler Data Joint Analysis Reveals that HD 175167b is a Massive Cold Jupiter}

\correspondingauthor{Tianjun Gan}
\email{tianjungan@gmail.com}

\author[0000-0002-4503-9705]{Tianjun~Gan}
\affil{Department of Astronomy, Tsinghua University, Beijing 100084, People's Republic of China}




\begin{abstract}

HD 175167b is a cold ($P_{b}\sim 1200$ days) Jupiter with a minimum mass of $M_{p}\sin i=7.8\pm3.5\ M_J$ orbiting a Sun-like star, first discovered by the Magellan Planet Search Program based on MIKE observations. Through a joint analysis of the MIKE data and the Gaia two-body orbital solution, \cite{Winn2022} found a companion mass of $M_{p}=14.8\pm1.8\ M_J$ and suggested that it might be better designated as a brown dwarf. Additional publicly available radial velocity data from Magellan/PFS better constrains the model, and reveals that the companion is a massive cold Jupiter with a mass of $M_p=10.2\pm0.4\ M_{J}$ and a period of $P_b=1275.8\pm0.4$ days. The planet orbit is inclined by $i=38.6\pm1.7^{\circ}$ with an eccentricity of $0.529\pm0.002$. 

\end{abstract}



\section{Introduction} \label{sec:intro}

HD 175167b is a long-period massive gas giant around a G5V star reported by \cite{Arriagada2010} using the MIKE radial velocity (RV) data taken under the Magellan Planet Search Program \citep{Bernstein2023}. The host star has a mass about $1\ M_{\odot}$ while the planet has a period of $1290\pm22$ days with a minimum mass of $M_{p}\sin i=7.8\pm3.5\ M_J$ on an eccentric ($e=0.54\pm0.09$) orbit. The minimum mass of the companion was later refined to $9.0\pm3.3\ M_J$ by \cite{Stassun2017} based on a better-constrained empirical stellar mass. More recently, \cite{Feng2022} and \cite{Xiao2023} both carried out full orbit fits with a combination of ground-based RV measurements and astrometric data from Hipparcos \citep{Perryman1997,vanLeeuwen2007} as well as Gaia \citep{Gaia2016,Gaia2018,Gaia2021}, enabling them to determine the true mass of the secondary. However, since the time-series Gaia astrometric data have not been published and only two absolute astrometry measurements are available now at the epoch of Year 1991 and 2016, the derived companion mass ($M_{p}=6.4\pm1.4\ M_J$ and $9.8\pm1.9\ M_J$) and other orbital parameters still have large uncertainties. 

As part of Gaia Data Release 3 \citep[Gaia DR3;][]{Gaia2023}, \cite{Halbwachs2023} presented a ``two-body'' catalog (\code{nss\_two\_body\_orbit}) based on Gaia observations spanning about 33 months. It includes orbital solutions for about 165,500 systems, among which 1162 sources are potential substellar objects \citep{Holl2023}. Each orbital solution contains the best-fit orbital parameters like four Thiele-Innes coefficients \citep{Binnendijk1960,Heintz1978}, period, eccentricity, time of periastron and parallax along with a correlation matrix between them. A recent work from \cite{Winn2022} proposed a method to conduct a joint analysis of these two-body solutions and RV data, which has been used in many relevant studies \citep[e.g.,][]{Unger2023,Fitzmaurice2023}. Basically, the Gaia two-body solution is treated as a basis and the correlation matrix is transformed to a covariance matrix using the \code{nsstools}\footnote{\url{https://www.cosmos.esa.int/web/gaia/dr3-nss-tools}} algorithm \citep{Halbwachs2023} to construct a Gaia likelihood function. MCMC sampling is then performed to maximize the combined Doppler$+$Gaia likelihood function. Note that this approach is only valid when the system contains a single companion. For multi-companion systems, the Gaia two-body solution from a single Keplerian model fit would not be robust thus the joint fit cannot be done. 

With this method, \cite{Winn2022} performed a joint analysis for the HD 175167 system using the published MIKE data and the Gaia two-body solution, leading to a companion mass of $14.8\pm1.8\ M_J$, a period of $1175\pm25$ days and an orbital inclination of $35.5\pm2.3^{\circ}$. Therefore, \cite{Winn2022} suggested that the companion might be located in the brown dwarf mass range \citep{Burrows1997,Spiegel2011}. However, two caveats were also pointed out by \cite{Winn2022}: (1) the Gaia orbital solution has large uncertainties, which may be due to the shorter time span of Gaia observations compared with the companion's period; (2) the periods and eccentricities are discrepant between the Gaia and Doppler solutions, which is perhaps caused by non-Gaussian uncertainties of Gaia orbital parameters.







\section{Magellan/PFS data and Joint Analysis}

In addition to the MIKE radial velocity data, HD 175167 was also monitored by the Planet Finder Spectrograph \citep[PFS;][]{Crane2006,Crane2008,Crane2010} on the 6.5 m Magellan II (Clay) telescope between 2010 May 23 and 2019 August 14 (see the data published in \citealt{Feng2022}). A total of 22 observations were collected with a median RV uncertainty of 1.1~m/s. 

\begin{figure*}
\centering
 \includegraphics[width=0.85\textwidth]{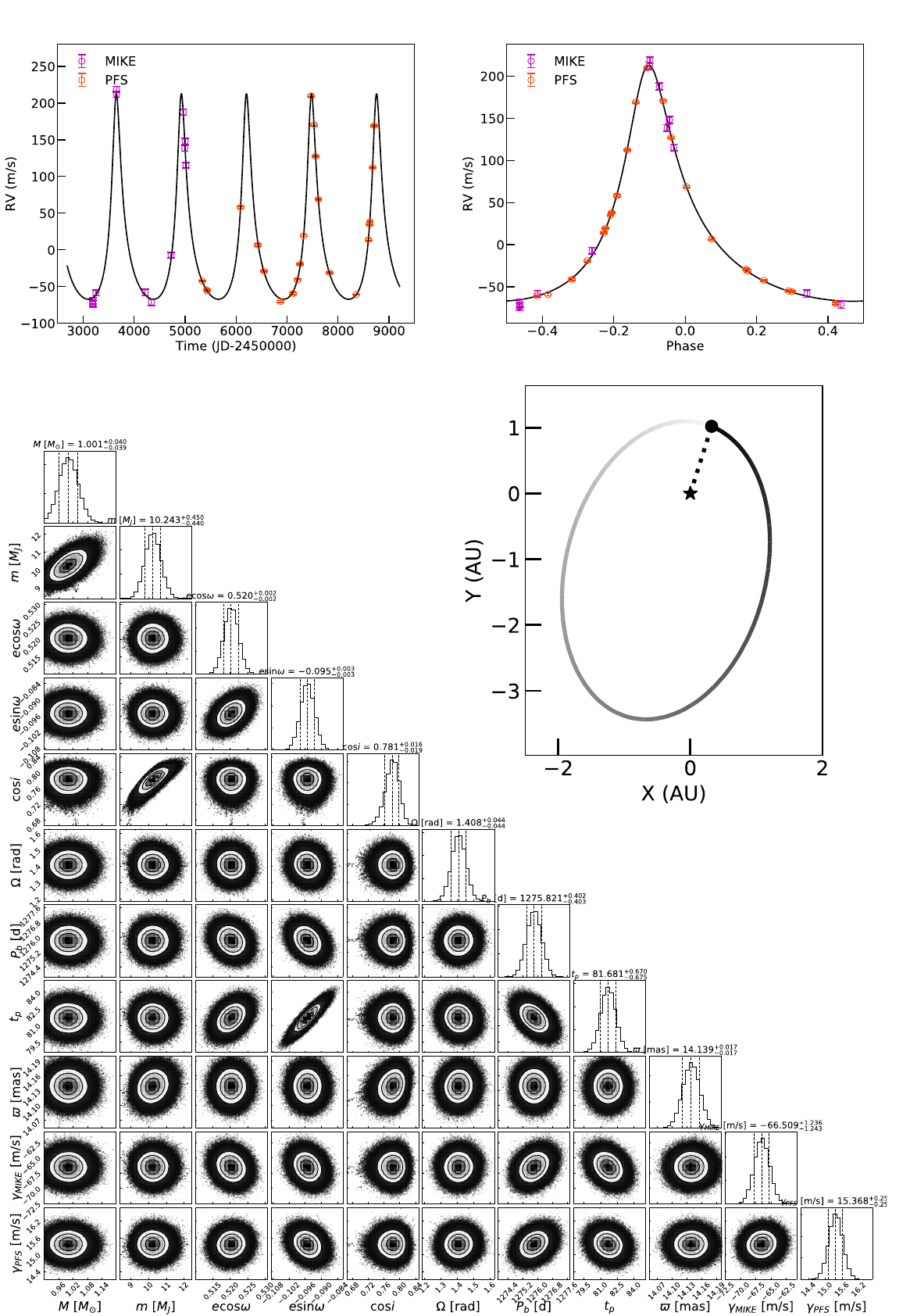}
 \caption{{\it Top panels}: Time-series (left) and phase-folded (right) RV data from MIKE and PFS along with the best-fit model. {\it Middle right panel}: The orbit of the HD 175167b constructed based on the best-fit physical parameters using \code{Rebound} \citep{rebound}. The parent star is located at the origin. {\it Bottom panel}: The posterior distributions of the joint-fit analysis. We follow the Gaia convention where $t_{p}$ is the Julian date of pericenter minus 2,457,389.}
 \label{HD175167}
\end{figure*}

Following the methodology proposed by \cite{Winn2022}, I perform a joint-fit analysis including the PFS data. The joint model has 11 free parameters including stellar mass $M$, companion mass $m$, $e\cos \omega$, $e\sin w$, the cosine of the inclination $\cos i$, the longitude of the ascending node $\Omega$, orbital period $P$, time of periastron $t_{p}$, parallax $\varpi$ and two RV offsets $\gamma_{\rm MIKE}$, $\gamma_{\rm PFS}$, where $e$ is the orbital eccentricity and $\omega$ represents the argument of pericenter. The Keplerian model is built by utilizing the code \code{radvel} \citep{Fulton2018}. Since the companion is a dark body based on its mass measurement from previous works, the flux ratio $\varepsilon$ between the companion and the primary star is thus fixed at 0. Except for a Gaussian prior $\mathcal{N}$ ($1.00$\ ,\ $0.04^{2}$) adopted on the host star mass based on the value from literature \citep{Feng2022}, all the other parameters are set with non-informative uniform priors. The posterior sampling is carried out with \code{emcee} \citep{Foreman2013}. A total of 100 walkers are initialized and all of them are run for 35,000 steps with the first 5,000 steps excluded as burnt-in samples. \autoref{HD175167} presents the final joint-fit results, the orbital geometry made by \code{Rebound} \citep{rebound} as well as the posterior distributions. 

\section{Results and Discussions}

After including additional RV data points, the uncertainty on the companion mass is successfully reduced. The joint-fit of Doppler velocities from Magellan/MIKE and Magellan/PFS along with the Gaia two-body solution reveals that the companion has a mass of $10.2\pm0.4\ M_J$, an eccentricity of $e=0.529\pm0.002$ and a period of $1275.8\pm0.4$ days. The planet orbital period constrained by the new joint-fit here is better in accord with the values $1290.0\pm22.0$, $1276.1\pm0.7$ and $1289^{+8}_{-28}$ days from \cite{Arriagada2010}, \cite{Feng2022} and \cite{Xiao2023}, compared with $1175\pm25$ days determined by \cite{Winn2022}. The best-fit planet orbit inclination is $38.6\pm 1.7^{\circ}$, which is consistent with the result $35.5\pm2.3^{\circ}$ reported by \cite{Winn2022} but different from $94.6\pm23.7^{\circ}$ and $60\pm17^{\circ}$ measured by \cite{Feng2022} and \cite{Xiao2023}. One possibility is that the small number (2 data points from Hipparcos and Gaia) of absolute astrometry measurements limits the accuracy and precision of the orbital inclination measurement in \cite{Feng2022} and \cite{Xiao2023}. Based on the updated mass measurement from the joint analysis, HD 175167b turns out to be a massive cold gas giant planet rather than a brown dwarf.


\section{Acknowledgments}

The author thanks Drs.~Shude Mao, Sharon X. Wang and Joshua Winn for the useful discussions. This work is supported by the National Science Foundation of China (Grant No. 12133005).


\facilities{Gaia, Magellan/MIKE, Magellan/PFS}

\software{radvel \citep{Fulton2018}, emcee \citep{Foreman2013}, Rebound \citep{rebound}}

\bibliography{planet}{}
\bibliographystyle{aasjournal}



\end{document}